\documentclass[preprint,review,12pt]{elsarticle}
\usepackage{amssymb}
\usepackage{lipsum}
\usepackage{multicol}
\usepackage{xcolor}

\journal{Physics Letters B}

\begin{document}

\begin{frontmatter}

\title{IceCube data for astrophysical neutrinos exhibit $q$-statistical behavior}

\author[1,2,3,4]{Constantino Tsallis}
\author[5,6]{Ugur Tirnakli}

\address[1]{Centro Brasileiro de Pesquisas Fisicas, Rua Xavier Sigaud 150, 22290-180, Rio de Janeiro RJ, Brazil}
\address[2]{Santa Fe Institute, 1399 Hyde Park Road, 87501 NM, Santa Fe, USA}
\address[3]{Complexity Science Hub, Metternichgasse 8, 1030, Vienna, Austria}
\address[4]{Dipartimento di Fisica e Astronomia Ettore Majorana, Via S. Sofia 64, 95123, Catania, Italy}
\address[5]{Department of Physics, Faculty of Arts and Sciences, Izmir University of Economics, 35330, Izmir, Turkey}
\address[6]{Complex Systems Research and Application Center, Izmir University of Economics, 35330, Izmir, Turkey}

\begin{abstract}
Relevant IceCube recent data for astrophysical neutrinos are analyzed. Similarly to cosmic rays and proton-proton LHC collisions, the corresponding spectra are satisfactorily described, for the entire interval of measured energies, through $q$-Gaussians. This fact reinforces the already available evidence that nonadditive entropies (either of the $S_q$ class or of the $S_\delta$ one, or even alternative ones) appear to play a crucial role in systems such as black holes and related cosmological phenomena (e.g., dark energy). The IceCube data yield high values for the momenta-distribution nonadditive index ($q \simeq 1.6$, hence $q_{energy} \simeq 1.3$), to be compared with the cosmic rays values ($q_{energy} \simeq 1.2$) and the LHC values ($q_{energy} \simeq 1.1$). This fact neatly indicates very-long-ranged space-time correlations. The detailed associated physical mechanisms remain, at the present stage, as an instigating open issue.
\end{abstract}




\end{frontmatter}




\section{Introduction}
\label{introduction}
\begin{figure}
\includegraphics[width=1.1\columnwidth]{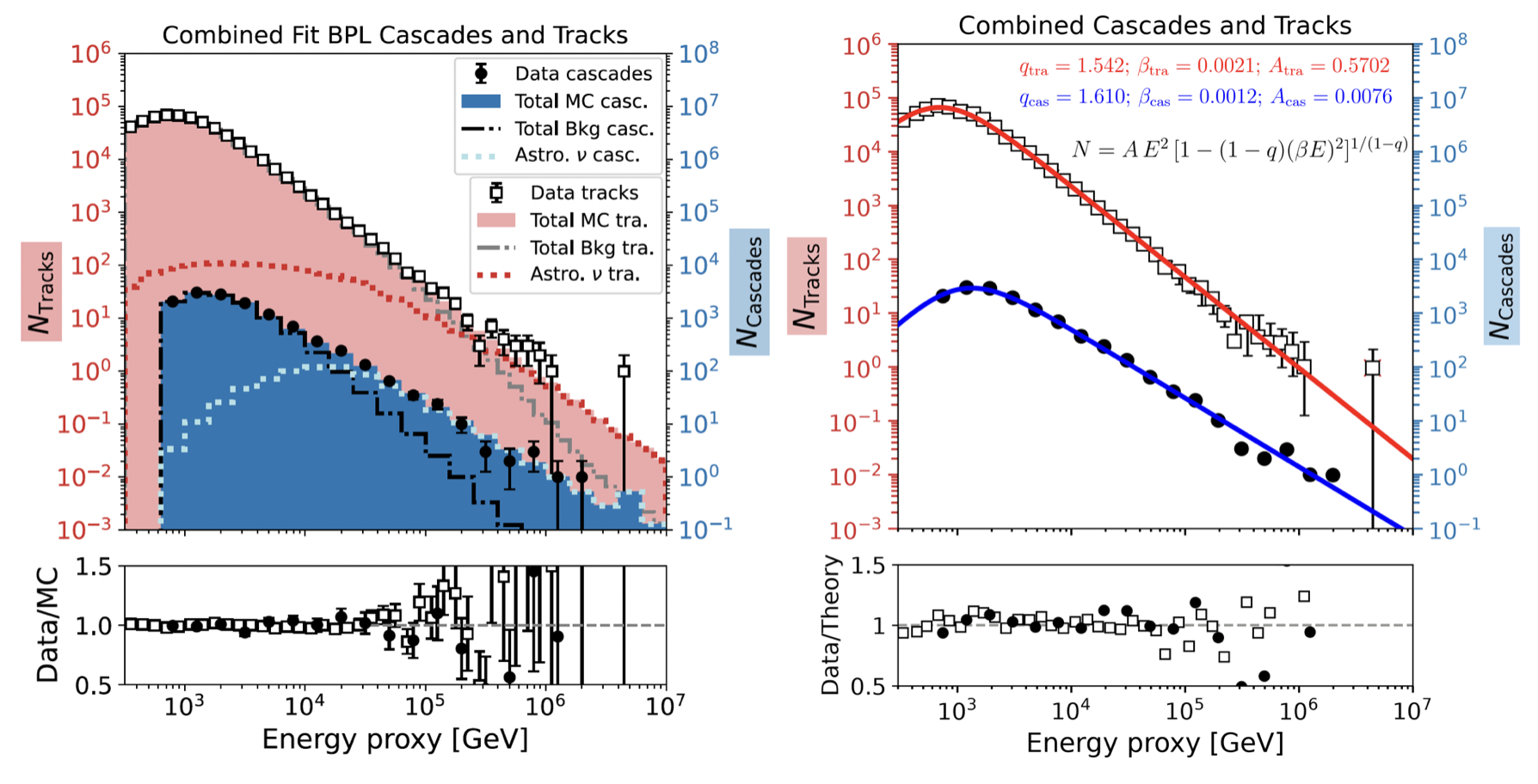}
\includegraphics[width=1.1\columnwidth]{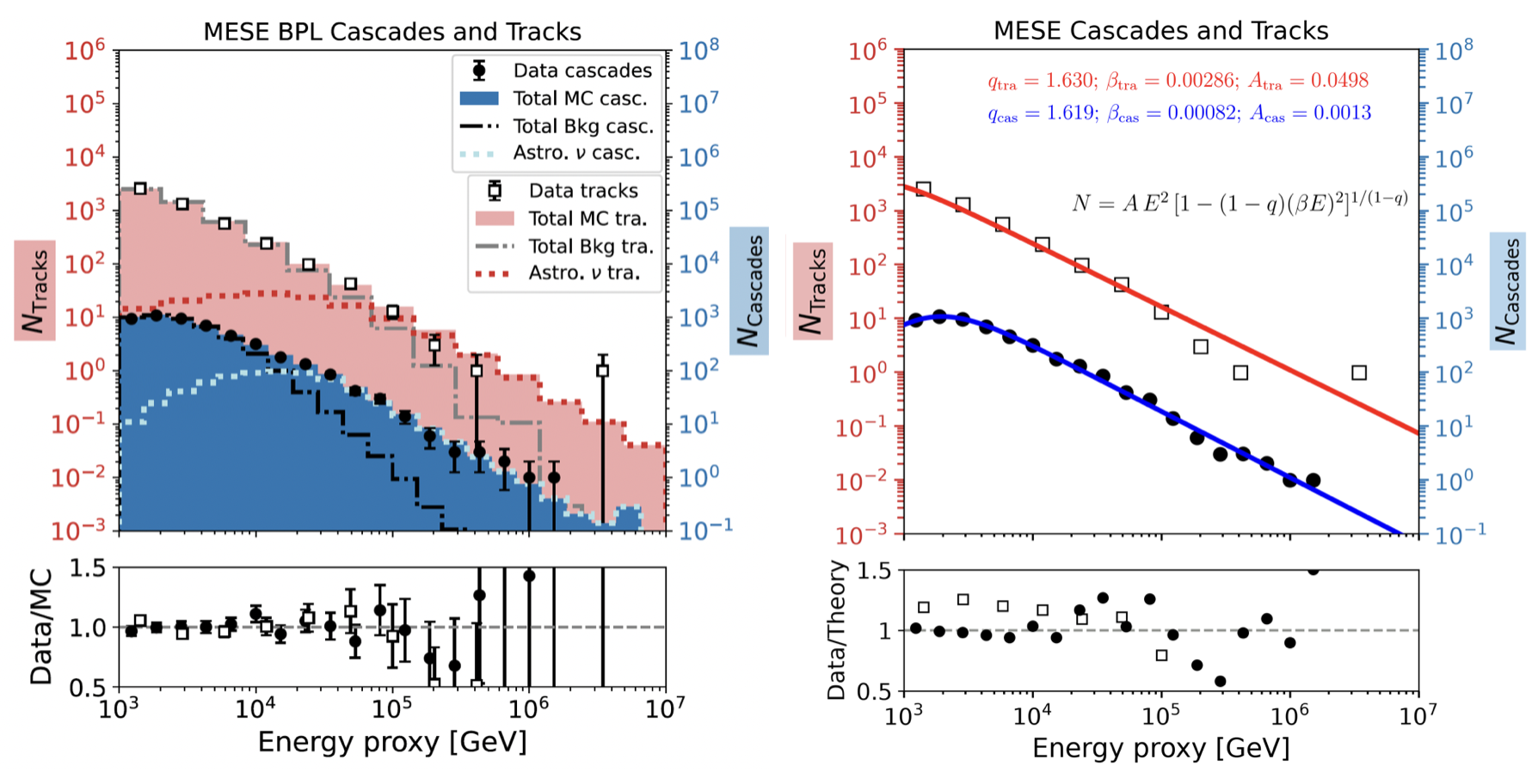}
  \caption{Fluxes for the neutrino energies (momenta) in the range TeV-PeV, for the CF (Combined Fit; top plots) and the MESE (Medium Energy Starting Events; bottom plots)
analyses. Left plots: from \cite{AbbasiPRL}. Right plots: present results.}
\label{figCombined}
\end{figure}

Remarkable observations of astrophysical neutrinos at the IceCube Observatory 
are presently available \cite{AbbasiPRL,AbbasiPRD}. The next natural scientific step is the attempt to deeply understand why they
exhibit the specific behavior that has been detected in those observations. Along this line, it is definitively helpful to have exact or approximate analytical expressions satisfactorily approaching the empirical data. Preliminary analytical expressions have already been advanced in \cite{AbbasiPRL,AbbasiPRD}, namely within the log-parabolic (LP) and broken power-law (BPL) models. The former appears to be appropriate for high energies but fails for low energy ones; the latter has the opposite tendency, being appropriate for low energies but failing for high ones (see the left plots of Fig.~\ref{figCombined}. The goal of the present Letter is to identify a single (and simple) analytical form, namely the $q$-Gaussian form, which is satisfactory for the {\it entire range of measured energies}. This form, which recovers the Gaussian one as its $q=1$ particular instance, not only satisfactorily fits, for $q>1$, all the data presented in Fig.~\ref{figCombined} but also provides {\it quantitative} evidence of an important clue for deeper understanding, namely that the relevant space-time correlations definitively are {\it long-ranged}, which naturally emerges as a most plausible fact.

It is well known that simple systems (e.g., air in laminar quasi-equilibrium regime, standard classical or quantum ferromagnets, BCS superconductors) involve short-range time-space correlations and are very satisfactorily described within the usual, more-than-centennial, Boltzmann-Gibbs (BG) statistical mechanics (\cite{Reichl2016} and references therein). 

In significant contrast, complex systems involve long-range correlations and require instead more powerful theoretical frames. Such systems are plethoric in nature (as well as in technological and social complex systems) and include long-range-interacting classical many-body systems \cite{ChristodoulidiTsallisBountis2014,CirtoRodriguezNobreTsallis2018}, overdamped many-body systems \cite{AndradeSilvaMoreiraNobreCurado2010,CasasNobreCurado2019}, granular matter \cite{CombeRichefeuStasiakAtman2015}, quantum-tunneling chemical reactions \cite{WildNotzoldSimpsonTranWester2023}, cold atoms in dissipative optical lattices \cite{DouglasBergaminiRenzoni2006,LutzRenzoni2013}, turbulence \cite{DanielsBeckBodenschatz2004,LimaPereiraMoriconiSreenivasanTsallis2026,BeckTsallis2026}, solar wind \cite{BurlagaVinas2005}, astrophysical objects and cosmological issues \cite{TsallisCirto2013,JizbaLambiase2022,SalehiPouraliAbedini2023,JizbaLambiase2023,DenkiewiczSalzanoDabrowski2023,TsallisJensen2025}, cosmic rays \cite{TsallisAnjosBorges2003}, high-energy collisions of elementary particles and heavy ions at LHC/CERN and PHENIX/Brookhaven \cite{PHENIX2011,WongWilkCirtoTsallis2015,DeppmanMegiasMenezes2020},
nonlinear dynamical systems at the edge of chaos \cite{BaldovinRobledo2004,TirnakliTsallisBeck2009,TirnakliBorges2016,SaberiTirnakliTsallis2025}, spin-glass \cite{PickupCywinskiPappasFaragoFouquet2009,OnderSaberiMoessner2026},
asymptotically scale-free networks \cite{ThurnerTsallis2005,CinardiRapisardaTsallis2020,OliveiraBritoSilvaTsallis2022},  financial systems \cite{Borland2002}.  A frequent and intensively studied paradigmatic path is constituted by  thermostatistical approaches based on {\it nonadditive} entropic functionals, such as $S_q$ \cite{Tsallis1988}, $S_\delta$ \cite{Tsallis2009}, $S_{q,\delta}$ \cite{TsallisCirto2013} and others (Renyi \cite{Renyi1961}, Kaniadakis \cite{Kaniadakis2001,Kaniadakis2002,Kaniadakis2005}, Hanel-Thurner \cite{HanelThurner2011a,HanelThurner2011b}, Tempesta \cite{Tempesta2020,TsallisJensen2025} entropic functionals). All of these recover the celebrated Boltzmann-Gibbs-von Neumann-Shannon {\it additive} entropic functional 
\begin{equation}S_{BG}(\{p_i\})=-k\sum_{i=1}^W p_i \ln p_i
\end{equation}
as their simplest particular instance (e.g., $q=\delta=1$). The overall scenario is tutorially focused on in \cite{Tsallis2023} and various related mathematical issues are discussed in \cite{UmarovTsallis2022}, the foundational assumption being that, once the set of $\{p_i\}$ is expressed as function of the size $N$ of the system we are focusing on, the entropy $S(N))$ has to be {\it extensive} (meaning that $S(N) \equiv S(\{p_i(N)\}) \propto N$ in the $N \to\infty$ limit), as mandated by the Legendre structure of classical thermodynamics (see, for instance, an early illustration introduced in \cite{TsallisGellMannSato2005}). 

Let us focus now on the IceCube data \cite{AbbasiPRL,AbbasiPRD}.

%
%
%
%
\begin{table}[htbp]
  \centering
  \caption{Parameter values for all cases.}
  \label{tab:4x3}
  \begin{tabular}{|c|c|c|c|}
    \hline
     & $A$ & $\beta$ & $q$ \\
    \hline
    Combined Tracks & $0.57\pm0.11$ & $0.00212\pm 0.00078$ & $1.5425\pm 0.0034$\\
    \hline
    Combined Cascades & $0.0076\pm 0.0030$ & $0.00119\pm 0.00066$ & $1.6103\pm 0.0074$\\
    \hline
    MESE Tracks & $0.05\pm 0.56$ & $0.0029 \pm 0.0077$ & $1.629 \pm 0.027$\\
    \hline
    MESE Cascades & $0.00132\pm 0.00044$ & $0.000824 \pm 0.000130$ & $1.6193\pm 0.0064$\\    \hline
  \end{tabular}
\end{table}


\section{The IceCube data and the $q$-thermostatistical approach}
In Figure 1 of \cite{AbbasiPRL}, we see the IceCube data corresponding to the energy distributions of astrophysical neutrinos for both Tracks and Cascades within the CF (Combined Fit) and the MESE (Medium Energy Starting Events) analyses. We also see, for each of these four data sets, two possible approaches associated with the log-parabolic (LP) and broken power-law (BPL) models. In all four cases, the LP model satisfactorily fits the data at high energies but fails at low energies, the BPL model corresponding to the other way around. Our present aim is to show that a simple functional form, namely the $q$-Gaussian distribution, analytically grounded on nonadditive entropies satisfactorily fits the data along the {\it entire energy range}.

The additive BG entropic functional (1) has been generalized \cite{Tsallis1988,Tsallis2023} into the nonadditive one
\begin{equation}
S_q({p_i})=k\sum_{i=1}^W p_i \ln_q \frac{1}{p_i} \;\;\; (q \in \mathbb{R})\,,
\end{equation}
where $\ln_z \equiv \frac{z^{1-q}-1}{1-q} \;\; (z>0;\, \ln_1 z=\ln z)$. The $q$-Gaussian distribution is obtained by optimizing, under a simple constraint such as (for $d=1$) $\int dx\,x^2 f(x)=constant$, the form $S_q=k\int dx f(x) [\ln_q 1/f(x)]$, $x$ being a dimensionless variable (see details in \cite{DogniniTsallis2026} and references therein). For $d$ dimensions, the $q$-Gaussian distribution is defined 
as follows:
\begin{equation}
f_q(x)=f_q^{0}\,x^{d-1} e_q^{-\bar\beta \,x^2} \equiv \frac{f_q^{0}\, x^{d-1}}{[1+(q-1)\bar\beta\,x^2]^{\frac{1}{q-1}}} \;[q \ge 1;\, f_1(x)=f_1^{0}x^{d-1}e^{-\bar\beta x^2}] \,,
\end{equation}
where $\bar\beta$ plays the role of an effective inverse temperature and $f_q^0$  is a normalizing factor (analytically obtained by imposing $\int_{0}^{\infty} dx\, x^{d-1}f_q(x)=1$). For future convenience (in order to comply with the same units used in \cite{AbbasiPRL}), let us introduce $\bar\beta \equiv \beta^2$. Therefore, the $q$-Gaussian we are going to use is given by
\begin{equation}
f_q(x)=f_q^{0}\,x^{d-1} e_q^{-(\beta \,x)^2} \, 
\end{equation} 

In our fittings, we have adopted (as already mentioned above) the same notation that is used in \cite{AbbasiPRL} and we have interpreted the data for the flux ($N$) as follows:
\begin{equation}
N=A\,E^2\,e_q^{-(\beta\,E)^2} \,.
\end{equation}
Notice that the neutrinos are in the ultrarelativistic regime; therefore, their energy is given by $E=\sqrt{p^2c^2+m^2c^4} \sim pc$. In other words, we are fitting with a $d=3$ $q$-Gaussian distribution of momenta $p \equiv |\vec{p}|$, namely of the form $N =a\,p^{d-1}\,e_q^{-(\beta\,p)^2}\;\; (a>0, b>0)$. Naturally, this distribution recovers, for $q=1$ (i.e., within the BG equilibrium thermostatistics), the celebrated Maxwellian distribution of $d$-dimensional momenta. Let us also mention that if we wish to asymptotically compare $e_q^{-(\beta\,p)^2} \sim 1/(p^2)^{1/(q-1)}$ with $e_{q_{energy}}^{-\beta_{energy}\,p} \sim 1/p^{1/(q_{energy}-1)}$, we need to identify
\begin{equation}
\frac{2}{q-1}=
\frac{1}{q_{energy}-1} \,.
\end{equation}
Let us finally mention at this point that, for particularly complex systems such as fully turbulent ones, functional forms that are more general than the present one, namely $N =a\,p^{d-1}\,e_q^{-(\beta\,p)^\delta}\;\; (\delta \ge 1)$, have been satisfactorily used \cite{LimaPereiraMoriconiSreenivasanTsallis2026,BeckTsallis2026}.  

\section{Conclusions}
Our fittings yield, in all the cases that have been focused on here, $q \simeq 1.6$, hence $q_{energy}\simeq 1.3$, with $1/\beta$ roughly in the range $10^6 \,GeV=PeV$. The values for $q_{energy}$ are larger than those that are typical for cosmic rays, namely $q_{energy}\simeq 1.2$ \cite{TsallisAnjosBorges2003}, or typical for proton-proton collisions at LHC/CERN, namely $q_{energy}\simeq 1.1$ \cite{WaltonRafelski2000,WongWilkCirtoTsallis2015,MegiasDeppmanPasechnikTsallis2023}. Summarizing, $q_{energy}^{IceCube} \simeq 1.3 > q_{energy}^{Cosmic Rays} \simeq 1.2 > q_{energy}^{LHC} \simeq 1.1 > q_{BG}=1$. 
These inequalities imply, for the present astrophysical neutrinos, that their travel from very distant sources to the IceCube detectors occurs in the presence of {\it very-long-ranged}  space-time correlations; indeed, the range of the correlations monotonically increases with the index $q$, the limiting value $q=1$ corresponding to short-range correlations. The precise nature of the very-long-ranged correlations of those astrophysical neutrinos  remains, though, as an instigating open problem. 

Last but not least, we incidentally verify (see the four narrow bottom graphs of Fig. \ref{figCombined}) that rough log-periodic oscillations are lurking, which is not a rare feature in such systems (see \cite{WongWilkCirtoTsallis2015, WilkWlodarczyk2014,WilkWlodarczyk2017}). This fact strongly suggests that some sort of self-similar structure (multifractal-like, for instance) has emerged.

\section*{Acknowledgements}
We thank fruitful comments from C. Beck, A.A. Saberi and R. Pasechnik. 
C.T. has benefited from partial financial support by CNPq and FAPERJ (Brazilian Agencies).
This study was supported by Scientific and
Technological Research Council of Türkiye (TUBITAK) under Grant Number 126F100. The authors thank TUBITAK for their support.
U.T. is a member of the Science Academy, Bilim Akademisi.




\end{document}